\documentclass[final,5p,times,twocolumn,numbers]{elsarticle}

\usepackage{amssymb}
\usepackage{lipsum}
\usepackage{xcolor}
\usepackage{framed}
\usepackage{paralist}
\usepackage{amsmath}
\usepackage{graphicx}
\usepackage{subcaption} 
\usepackage{float}
\usepackage{array}
\usepackage{multirow}
\usepackage{booktabs} 

\begin{document}
\begin{frontmatter}

\newcommand\Ren{\mbox{\textit{Re}}}  
\newcommand\Ra{\mbox{\textit{Ra}}}
\newcommand\Cn{\mbox{\textit{Cn}}}
\newcommand\Ca{\mbox{\textit{Ca}}}
\newcommand\We{\mbox{\textit{We}}}
\newcommand\Fr{\mbox{\textit{Fr}}}
\newcommand\Bo{\mbox{\textit{Bo}}}
\newcommand\Nun{\mbox{\textit{Nu}}}
\newcommand\Prn{\mbox{\textit{Pr}}} 
\newcommand\Pe{\mbox{\textit{Pe}}}  

\title{Numerical study of freezing efficiency for a moving  droplet in the microchannel}

\author[1]{Dahu Li}

\author[1]{Z. L. Wang \corref{cor1}}
\ead{wng\_zh@i.shu.edu.cn}

\address[1]{Shanghai Institute of Applied Mathematics and Mechanics, School of Mechanics and Engineering Science, Shanghai University, Shanghai, 200072, China}

\cortext[cor1]{Corresponding author}

\begin{abstract}

In microfluidic devices, droplets serving as carriers for chemical reactors or biomass can form stably encapsulated particles during the freezing process, holding significant importance in pharmaceuticals and microchemical reaction control. The droplet freezing process involves multiple complex stages, including supercooling, recalescence, solidification, and cooling. Current research in microfluidics within this field still has limitations, necessitating deeper and more systematic exploration. This study couples the Volume of Fluid (VOF) interface model with an enthalpy-porous media phase change model to distinguish the water (ice)-oil two-phase system and the water-ice two-phase system, respectively. Through numerical simulation of the liquid-liquid-solid (oil-water-ice) three-phase motion behavior, we aim to reveal the solidification patterns of moving droplets under supercooled conditions within microchannels and the correlations related to freezing time efficiency. Droplets of different sizes, velocities, and under varying temperature difference conditions are investigated. The simulation reveals two distinct solidification patterns: Pattern I exhibits a uniform and symmetric pattern, occurring at lower Reynolds numbers ($Re$) or droplet-to-channel diameter ratios ($D/W$), resulting in a relatively even solid shell along the interface with synchronized solidification fronts in both flow and spanwise directions, dominated by heat conduction.  Pattern II shows a shear-constraint cooperative non-uniform pattern at higher Reynolds numbers  and $D/W$ ratios, where flow dynamics and spatial confinement couple, leading to faster solidification at the tail region and an asymmetric solidification front. Through comprehensive analysis of the numerical results, we derived the characteristic behavior of the Stefan number ($Ste$), Reynolds number ($Re$), and $D/W$ ratios on the freezing time. We established a unified scaling relation for the freezing time: $ t_\text{final}\sim18.02 S t e^{-0.91}{ Re }^{-0.12} (D / W)^{1.42} $. The results demonstrates that increasing Stefan number and Reynolds number shortens the freezing time, whereas increasing droplet $D/W$ ratio extends it. Furthermore, the fitted coefficients reveals that temperature  and droplet size exert a more pronounced influence on droplet freezing. Our results demonstrate good consistency with those in the literature, while indeed incorporating the effects of motion, which represents a novel aspect. This study elucidates the relevant mechanisms governing droplet freezing under motion within microscale effects, providing a theoretical basis for the design and control of micro-containers in microfluidic systems.

\end{abstract}


\end{frontmatter}

\section{Introduction}\label{Introduction}

\begin{table}[!ht]
    \centering
    \begin{tabular}{|p{15pt}p{200pt}|} 
        \hline 
        \multicolumn{2}{|l|}{\textbf{Nomenclature}} \\[5pt]
        \multicolumn{2}{|l|}{\textbf{Symbols}} \\[5pt]
      $Ste$& Stefan number  \\[0pt]
      $Re$ & Reynolds number \\[0pt]
      $Pr$ & Prandtl number \\[0pt]
      $We$ & Weber number \\[0pt]
      $L$ & Length of the channel ($\mathrm{mm}$) \\[0pt]
      $W$ & Width of the channel ($\mathrm{mm}$) \\[0pt]
      $D$ & Diameter of the droplet ($\mathrm{mm}$) \\[0pt]
      $t$ & Time ($\mathrm{ms}$) \\[0pt]
      $g$& Gravitational acceleration ($\mathrm{m/s^{2}}$) \\[0pt]
      $T$ & Temperature $({ }^{\circ} \mathrm{C})$ \\[0pt]
      $T_\text{wall}$ &  Wall temperature $({ }^{\circ} \mathrm{C})$ \\[0pt]
      $T_\text{sat}$ & Saturation temperature $({ }^{\circ} \mathrm{C})$ \\[0pt]
      $\Delta{T}$ & Temperature difference between the droplet and oil$({ }^{\circ} \mathrm{C})$ \\[0pt]
      $k$& Thermal diffusivity ($\mathrm{m^2/s}$) \\[0pt]
      $H$ & Enthalpy ($\mathrm{J/kg}$) \\[0pt]
      $h$  & Sensible enthalpy ($\mathrm{J/kg}$) \\[0pt]
      $\Delta H$& Latent heat  ($\mathrm{J/kg}$) \\[0pt]
      $l_\text{latent}$ &Phase change latent heat  ($\mathrm{J/kg}$) \\[0pt]
      $F_\text{S}$& Surface tension term \\[0pt]
      $S_\text{M}$& Momentum source term \\[0pt]
      $S_\text{E}$ &Energy source term  \\[0pt]
      $A_\text{mush}$ &Mushy zone constant \\[0pt]
      $u$& velocity($\mathrm{m/s}$) \\[0pt]
      $c_{p}$& Specific heat capacity($\mathrm{J/{kg\cdot K}}$) \\[5pt]
    \multicolumn{2}{|l|}{\textbf{Greek letters}} \\[5pt]
    $\alpha$ &Thermal expansion coefficient ($\mathrm{1/K}$) \\[0pt]
    $\gamma_\text{liquid}$ & Liquid volume fraction \\[0pt]
    $\sigma$&Surface tension ($\mathrm{N/m}$) \\[0pt]
    $\rho$ &Density ($\mathrm{kg/m^3}$) \\[0pt]
    $\mu$ &Viscosity of liquid $( \mathrm{Pa} \cdot \mathrm{s})$ \\[0pt]
    $\kappa$  &Interface curvature \\[0pt]
    $\varphi$& Volume fraction  of phase \\[0pt]
    $\epsilon$ &A small number  \\[5pt]
  \multicolumn{2}{|l|}{\textbf{Subscripts}} \\[5pt]
   w& water \\[0pt]
   o& oil \\[0pt]
    \hline 
    \end{tabular}
    \label{tab:one_column_table}
\end{table}

Droplet freezing is a phenomenon that is widely present in nature and holds significant importance in many fields of engineering and technology. Under cold climatic conditions, the freezing of tiny droplets in clouds and fog can trigger complex weather processes such as snowfall and freezing rain, which can have far-reaching impacts on transportation, agricultural production, and energy supply \cite{tabazadeh2002surface,demott2010predicting,mohler2007microbiology,cantrell2005production,zerr1997freezing}. In the field of aerospace, the rapid freezing of droplets on the wing surface when an aircraft passes through clouds can damage the aerodynamic shape, significantly increasing flight resistance and posing a serious threat to flight safety \cite{bragg2005iced,potapczuk2013aircraft,zhou2023review}. This also makes anti-freezing technology for aircraft a key area of research \cite{goraj2004overview,varanasi2010frost,wang2017recent,cui2024integrated}. In terms of industrial applications, in some projects involving low temperature environments, such as medicine and food cryopreservation \cite{li2002novel,ge2024research,zalazar2024situ},  the law of droplet freezing is directly related to the operation efficiency, energy consumption and product quality control of equipment.

The droplet freezing process encompasses four stages including supercooling, recalescence, solidification, and cooling \cite{song2020review,zhang2017freezing}. The duration of the recalescence stage is significantly shorter than that of the solidification stage \cite{jung2012mechanism}. In recent years, freezing of stationary droplets has been extensively and thoroughly investigated \cite{song2020review, zhao2020review, akhtar2023comprehensive, tiwari2023droplet}, encompassing both experimental and numerical studies on droplets stationary on cold substrates and suspended droplets \cite{chaudhary2014freezing, zhang2017freezing, zhang2017modelling, strub2003experimental, meng2020dynamic, meng2022freezing, zhang2018simulation}.The freezing of dynamic droplets is relatively less studied than that of static droplets. Sultana et al.\cite{sultana2017phase} used the Navi-er-Stokes equation coupled with the Volume of Fluid (VOF) method to trace the droplet-air interface, and numerically analyzed the phase transition of a free-falling droplet below 0 °C, and found that the internal circulation of the droplet enhanced the heat transfer and nucleation between the droplet and the surrounding environment, and the nucleation temperature of the large droplet is higher than that of the small droplet. Dong et al. \cite{zhao2017numerical} also studied a free falling droplet by using the lattice Boltzmann method (LBM), where the lower surface of the droplet is frozen first due to the strong cooling action of the air at the lower position and the influence of the flow field, however, for the whole droplet, freezing first occurred at the surface.

With the development of microfluidic technology, there are some experimental studies of microchannel droplet freezing reported. Sgro et al. \cite{sgro2007thermoelectric} were early to place Peltier elements either on top of or beneath the chip to freeze droplets. They utilized this device to freeze droplets for cell encapsulation and freezing, with applications in the field of biology. Stan et al. \cite{stan2009microfluidic} developed a high-precision microfluidic instrument that measures the ice nucleation temperatures of tens of thousands of water droplets within minutes with an accuracy of 0.4°C, and reveals the freezing differences between pure water and silver iodide (AgI)-seeded water. Tarn et al. \cite{tarn2021homogeneous,tarn2020chip} designed a microfluidic chip with two independent channel structures to generate water-in-oil droplets, and studied the uniform nucleation under the influence of the low temperature environment outside the channelline during the droplet movement, and calculated the uniform nucleation rate coefficient to verify the usability of the device.  Isenrich et al. \cite{isenrich2022microfluidic} generated droplets using a PDMS microfluidic platform, realized the continuous flow freezing of droplets in channellines using ethanol baths, and simulated the formation of ice nuclei in the atmosphere by studying the nucleation of pure water and microcline feldspar-containing suspensions. Shardt et al. \cite{shardt2022homogeneous} precisely quantified the homogeneous nucleation rate of water using an improved microfluidic device, reducing the temperature measurement uncertainty to 0.2 K. They obtained reliable data in the temperature range of 236.5–239.5 K (particularly above 238 K), addressing the large discrepancies in previous results within this range. Deck et al. \cite{deck2022stochastic} developed an open-source model tool that reveals the stochastic ice nucleation-dominated freezing process, validated with COVID-19 vaccine data, providing the basis for more rational tray freezing design.

Overall, research on the dynamic freezing processes of moving droplets remains very limited compared to studies involving stationary droplets. While some experimental reports on freezing moving droplets exist to date, they have primarily focused on droplet freezing nucleation. Due to the challenges in experimental observation, the ice front propagation during droplet dynamic freezing has not been thoroughly investigated, and numerical simulation studies on the freezing of moving droplets are even more scarce.

Therefore, we focus on microchannels to numerically investigate the ice front propagation during dynamic freezing of the moving droplet while evaluating their freezing efficiency, aiming to provide theoretical foundations for droplet freezing design in microfluidic systems. A two-dimensional axisymmetric physical and numerical model was established, coupling the Volume of Fluid (VOF) method with the enthalpy-porosity approach to simulate solidification phase change of moving droplets in the supercooled microchannel. Section \ref{Physical models and numerical methods} introduces the physical model and numerical methods. Section \ref{Result and discussion} conducts a grid independence study, describes the phenomenon of droplet freezing, discusses freezing patterns and mechanisms, and establishes scaling relations concerning droplet freezing time. The conclusion of this work is presented in section \ref{Conclusion}.

\section{Physical models and numerical methods}\label{Physical models and numerical methods}
\subsection{Physical models}

\begin{figure}[ht]
   \centering
    \includegraphics[width=8cm]{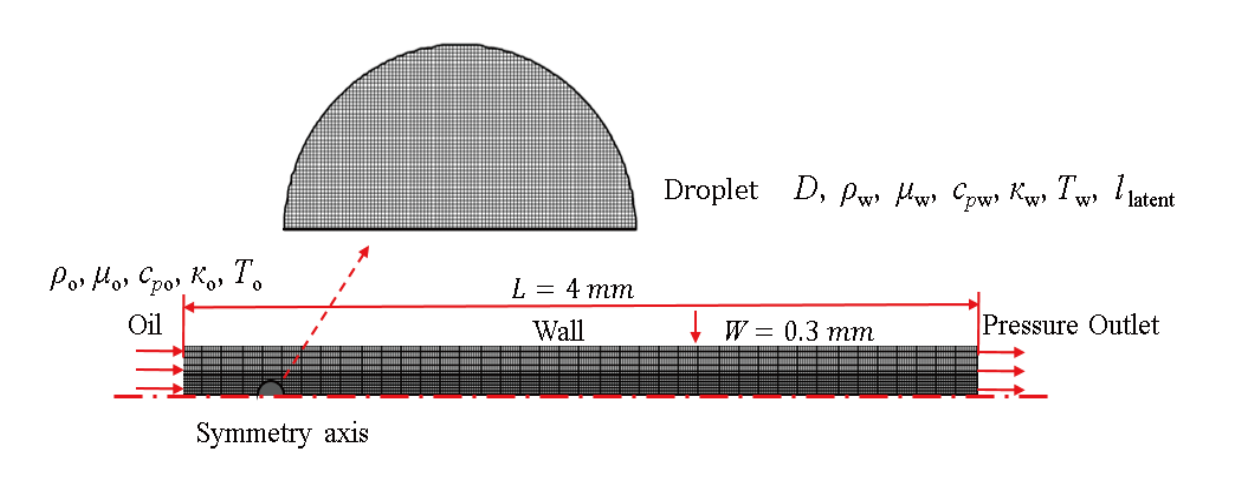}
    \caption{Physical mode of a moving droplet freezing in  microchannel}
  \label{Physical models}
\end{figure}

Figure \ref{Physical models} presents the physical model of droplet freezing during flow within a microchannel. The channel has a length of 4 mm and a diameter of 0.3 mm, with the left side serving as the inlet and the right side as the outlet. The upper and lower boundaries represent the channel walls. A droplet with a diameter of 0.1 mm is positioned within the inlet section. And the material parameters employed in simulations are listed in Table \ref{material}. Where, $D$, $\rho _\text{w}$, $\mu_\text{w}$, $ c_{p\text{w}}$, $k_\text{w}$, $T_\text{w}$ and $l_\text{latent}$ represent the droplet's diameter, density, dynamic viscosity, specific heat capacity, thermal conductivity, temperature, and latent heat of phase change, respectively;  $\rho _\text{o}$, $\mu_\text{o}$, $ c_{p\text{o}}$, $k_\text{o}$ and $T_\text{o}$ denote the density, dynamic viscosity, specific heat capacity, thermal conductivity, and temperature of the oil phase.

\begin{table*}[!htbp]
 \centering
 \caption{Material parameters (Employing fixed material parameters and disregarding temperature-dependent property variations at the microscale, testing confirmed their impact on droplet freezing to be below 2\%.)}
 \label{tab:pagenum}
 \begin{tabular}{cccccccc}
  \toprule
  \multirow{2}{*}{Material} & Specific heat  & Thermal conductivity & Density &   Viscosity &  Latent heat\\
   & $\left[\mathrm{J}/\left(\mathrm{kg} \cdot \mathrm{K}\right)\right]$& $[\mathrm{W} /(\mathrm{m} \cdot \mathrm{K})]$ & $\left(\mathrm{kg} / \mathrm{m}^{3}\right)$ & $( \mathrm{Pa} \cdot \mathrm{s})$  & $(\mathrm{kJ} / \mathrm{kg})$\\
  \midrule
  oil   & 2041.5 & 0.2& 851.45 & 0.004 & ---\\
  \midrule
     water    & 4220 & 0.6& \multirow{2}{*}{998.2}& 0.001 &\multirow{2}{*}{333.4}\\

    ice    & 2100 & 2.2&  & ---  & \\
  \bottomrule
 \end{tabular}
 \label{material}
\end{table*}

\subsection{Control Equation}
To analyze the performance of the fluid flow system, the following assumptions were made: (1) the liquid is considered a Newtonian fluid; (2) the fluid flow is incompressible, unsteady, viscous, laminar, and follows the Navier-Stokes equations; (3) the continuum hypothesis holds; (4) volume expansion during phase change is not considered. Therefore, mass conservation can be expressed as:
\begin{equation}
     \nabla \cdot \mathbf{u} = 0
\end{equation}
According to the laminar and unsteady flow characteristics of Newtonian fluids, the momentum equation can be obtained:
\begin{equation}   
    \centering
    \begin{split}
    \frac{\partial}{\partial t}(\rho \mathbf{u})+\nabla \cdot(\rho \mathbf{u} \mathbf{u})=-\nabla p+\nabla \cdot(\mu \nabla \mathbf{u})
    +\rho\mathbf{g}+\mathbf{F}_{\mathrm{s}}+\mathbf{S}_{\mathrm{M}}
    \end{split}
\end{equation}
where $\mathbf{F}_{\mathrm{s}}$ is the surface tension term, $\mathbf{S}_{\mathrm{M}}$ is the momentum source term for solidification, and the energy equation in the solidification and melting numerical problems is usually expressed in the following form:
\begin{equation}
    \centering
    \frac{\partial(\rho H)}{\partial t}+\nabla \cdot(\rho \mathbf{u} H)=\nabla \cdot(k \ \nabla {T})+S_\mathrm{E}
\end{equation}
where, $S_\mathrm{E}$ is the energy source term.

\subsubsection{Volume Of Fluid Model}
In the VOF model, different fluid components within the computational domain share a common set of momentum conservation equations during the simulation. The symbol $\varphi$ is defined to distinguish between the oil phase and the water (ice) phase:

\begin{equation}
\varphi = \begin{cases} 
1 & \text{Only oil phase in the control volume} \\
 0 \sim 1 & \text{Both oil and water (ice) phases in the control volume} \\
0 & \text{Only water (ice) phase  in the control volume}
\end{cases}
\end{equation}
The continuity equation can be expressed as:
\begin{equation}
    \frac{\partial \varphi}{\partial t}+\nabla \cdot(\mathbf{u} \varphi)=0 
\end{equation}

 The properties including the density, Viscosity  and thermal conductivity  are calculated from:
\begin{equation}
    \rho(\varphi)=\rho_\text{w} \varphi+\rho_\text{o}(1-\varphi)
\end{equation}

\begin{equation}
    \mu(\varphi)=\mu_\text{w} \varphi+\mu_\text{o}(1-\varphi)
\end{equation}

\begin{equation}
    k(\varphi)=k_\text{w} \varphi+k_\text{o}(1-\varphi)
\end{equation}

The Continuum Surface Force (CSF) model proposed by Brackbill et al.\cite{hirt1981volume} is used to calculate the interaction between two-phase fluids, and the volume force source term introduced into the momentum equation is:
\begin{equation}
    \mathbf{F}_{\mathrm{s}}=\sigma \frac{\rho \kappa \nabla \varphi}{\left(\rho_{o}+\rho_{w}\right) / 2} 
\end{equation}
where, $\kappa=\nabla \cdot \mathbf{n}$ is the curvature of the interface and $\mathbf{n}=\nabla \varphi /\left|\nabla \varphi\right|$ is the unit normal vector.

\subsubsection{Solidification/Melting model}

The Solidification/Melting model uses enthalpy-porosity technology to deal with solidification and melting problems that occur at isothermal or at a certain temperature, and determines the liquid fraction of the fluid by calculating the enthalpy equilibrium within each unit, and the region with the liquid fraction between $0\sim 1$  is the mushy zone\cite{voller1987fixed}.

In the Solidification/Melting model, the total enthalpy of a material is the sum of the sensible enthalpy and the latent heat of the solid-liquid mixture at a certain liquid fraction, that 
\begin{equation}
    H=h+\Delta H
\end{equation}
where $H$ is the total enthalpy, $h$ is sensible enthalpy and $\Delta H$ is the latent heat of the solid-liquid mixture.Where, the sensible enthalpy is defined as:
\begin{equation}
    h=h_{\mathrm{ref}}+\int_{T_{\mathrm{ref}}}^{T} c_p \mathrm{~d} T
\end{equation}
where $h_{\mathrm{ref}}$ is the apparent enthalpy of the reference state, $T_{\mathrm{ref}}$ is the temperature of the reference state, and $c_p$ is the specific heat. The latent heat of a solid-liquid mixture at a certain liquid fraction is defined as:
\begin{equation}
   \Delta H=\gamma_{\text {liquid }} l_{\text {latent}}
\end{equation}
where $l_{\text {latent}}$ is the latent heat of the phase change, $\gamma_{\text {liquid }}$  is the liquid fraction in the phase change region, which is defined as:
\begin{equation}
    \gamma_{\text {liquid }}=\left\{\begin{array}{cc}
0 & T \leq T_{\text {solid }} \\
\frac{T-T_{\text {solid }}}{T_{\text {liquid }}-T_{\text {solid }}} & T_{\text {solid }}<T<T_{\text {liquid }} \\
1 & T \geq T_{\text {liquid }}
\end{array}\right.
\end{equation}

In addition, the Solidification/Melting model treats the phase transition region as a porous medium, and the flow inside it satisfies Darcy's law of flow in a porous medium and conforms to the Carman-Koseny hypothesis, so the phase transition process introduces the source term of the momentum equation as\cite{voller1987fixed}:
\begin{equation}
    \mathbf{S}_{\mathrm{M}}=\frac{\left(1-\gamma_{\text {liquid }}\right)^{2}}{\gamma_{\text {liquid }}^{3}+\varepsilon} A_{\text {mush }} \mathbf{u}
\end{equation}
where $\varepsilon$ is a small number(0.001) introduced to prevent division by zero when  $\gamma_{\text{liquid}} = 0$ , $ A_{\text{mush}} $ is the mushy zone constant, set to 10000.

\subsection{Simulation setup}

The simulations are performed with the commercial CFD tool Ansys
 Fluent V.20.0. The process of droplet flow and freezing within the microchannel is unsteady, so it is necessary to give appropriate initial and boundary conditions. At the initial moment of calculation, the flow field is stationary and the microchannel is filled with oil at -35℃. To rapidly reduce the temperature of the droplet to the freezing point, a droplet with a temperature of 0.1 ℃ is patched near the inlet. The inlet is set with a parabolic velocity boundary condition for $u=u_{\max }\left(1-\frac{y^{2}}{({W/2})^{2}}\right)$ to flow oil  at a temperature of -35℃. The upper and lower boundaries of the channel are configured as  walls with no-slip  boundary conditions, and the outlet boundary condition is set as a pressure outlet. A pressure-based solver is adopted in the simulation, in which the Pressure-Implicit with Splitting of Operators (PISO) algorithm is used to couple pressure and velocity. Momentum, and energy governing equations are discretized using the second-order windward method, and the volume fraction is calculated by the Geo-Reconstruct method. The calculated material parameters are summarized in table \ref{table1}. In order to clearly capture the phase transition of the droplets in motion, the time step is set to 5e-7 s.

\section{Result and discussion}\label{Result and discussion}
\subsection{Grid Independence Verification}

\begin{figure}[!htbp]
    \vspace{-10pt}
    \centering
    \includegraphics[width=\linewidth]{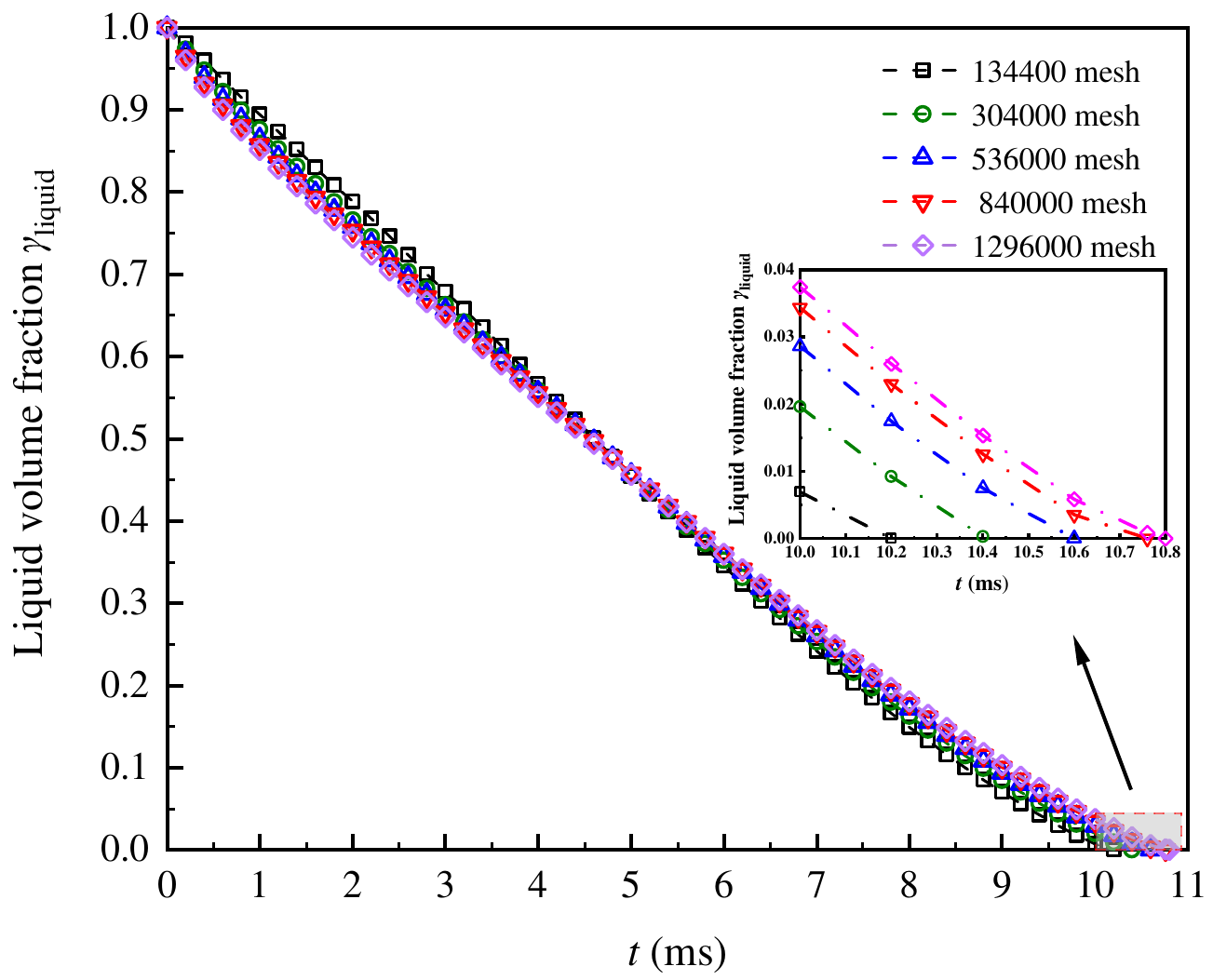}
    \caption{Temporal evolution of liquid volume fraction during droplet freezing under varying mesh resolutions}
    \label{Grid independence verification}
\end{figure}

To ensure that the numerical results are unaffected by grid resolution, five grid systems with mesh counts of 134,400; 304,000; 536,000; 840,000; and 1,296,000 are employed to demonstrate the temporal variation of liquid phase volume fraction during droplet freezing under the computational conditions of $Re = 10$ and $D/W = 0.33$, as shown in Fig.\ref{Grid independence verification}. Where $Re$ is the Reynolds number of the droplet ($Re=\frac{\rho_\text{w} u_{max} D}{\mu_\text{w}}$), and $D/W$ is the ratio of the droplet diameter to the Channel width. At the same time, the final freezing time of the droplet and the error with the densest number of grids are recorded in the table \ref{table1}. The simulation results are nearly identical for the five different grids systems, suggesting that the results are not affected by meshes. Considering the calculation accuracy and computational efficiency, the grid number of 840,000 is selected for the subsequent  simulation calculations.

\begin{figure*}[t]
   \centering
    \includegraphics[width=16cm]{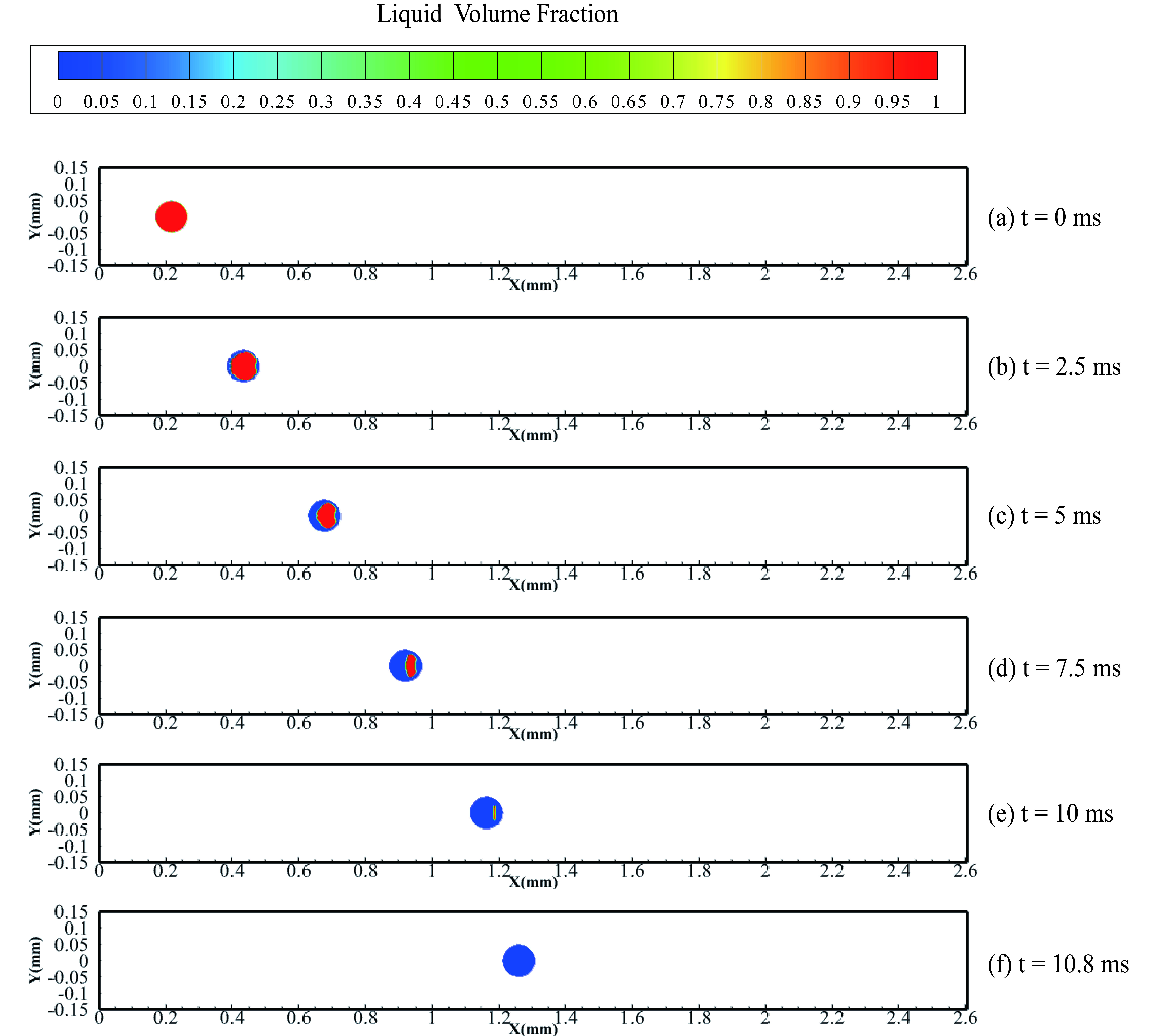}
    \caption{The Liquid volume fraction distribution of the droplet at different moments when the droplet flows within the microchannel.  The red area is the liquid region of the droplet freezing process ($\gamma_{liquid} = 1$), the blue area is the solid region of the droplet freezing process($\gamma_{liquid} = 0$), and the color transition area is the mushy region of the droplet freezing process($0 < \gamma_{liquid} < 1$).Peripheral freezing initiated at the droplet's outer surface. With progressive solidification, asymmetric ice formation developed. A distinct concave depression was observed at $t = 5$ ms, which gradually diminished by $t = 7.5$ ms. Complete solidification occurred at $t = 10.8$ ms.}
  \label{liquid volume fraction feild}
\end{figure*}

\begin{table}[H] 
\centering
\caption{ Grid dependence of droplet final freezing time}
\label{table1}
\begin{tabular}{cccc}
\hline
      &  Mesh number & Frezezing time (ms) & Error (\%)  \\
case1 &  134400   & 10.22     &   5.37\% \\
case2 &  304000    &  10.44      &    3.33\%  \\
case3 &  536000  &  10.62       &  1.67\%    \\
case4 &  840000    &    10.76     &   0.37\% \\
case5 &  1296000  &   10.80     &   ---\\

\hline
\end{tabular}
\end{table}

\begin{figure*}[t]
   \centering
    \includegraphics[width=16cm]{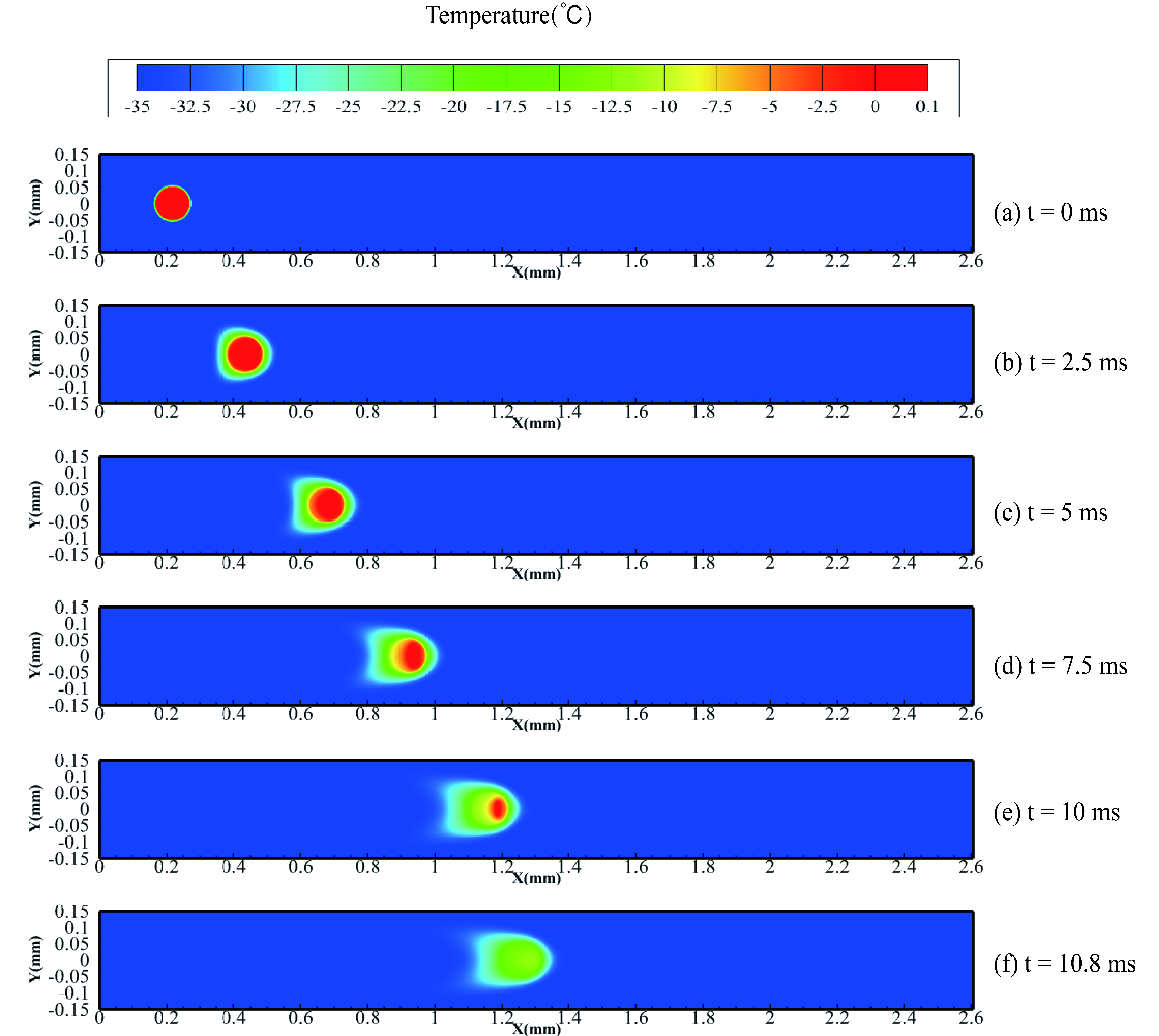}
    \caption{The temperature distribution of the droplet at different moments when the droplet flows within the microchannel.At $t= 0$ ms,the droplet and its surrounding temperature field exhibit a circular profile. As flow develops, the droplet evolves into a bullet-shaped morphology by $t=5$ ms. Concurrently, the internal droplet temperature decreases progressively, ultimately reaching a uniformly subzero state at $t=10.8$ ms.}
  \label{Temperature feild}
\end{figure*}

\begin{figure*}[!t]
   \centering
    \includegraphics[width=16cm]{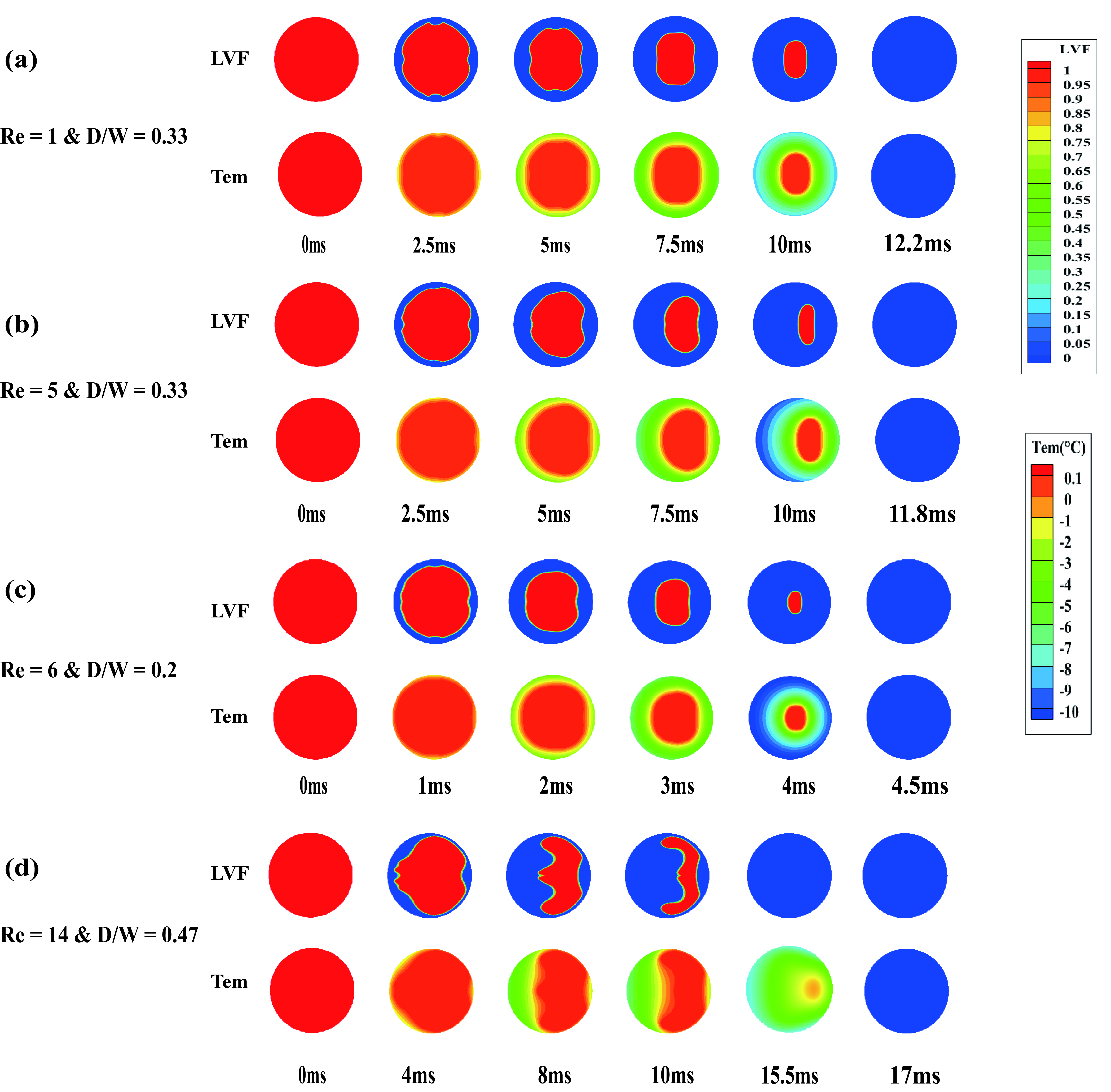}
    \caption{Droplet  Liquid volume fraction (LVF) distribution  contour and temperature (Tem) distribution  contour with different Reynolds numbers and different $ D/W $ numbers at different time: (a) $Re=1$ and $D/W=0.33$; (b)$Re=5$ and $ D/W=0.33$; (c) $Re = 10$ and $D/W = 0.33$; (d) $Re = 6$ and $D/W = 0.2$; (e) $Re = 14$ and $D/W = 0.47$.In computational cases (a) and (b), the droplet size remains fixed whereas the flow velocity is systematically varied. Conversely, in cases (c) and (d), a constant flow velocity is maintained while only the droplet size is modified.}
  \label{liquid volume fraction and temperature contour}
\end{figure*}

\subsection{Process of the droplet freezing}
Fig.\ref{liquid volume fraction feild} and \ref{Temperature feild} depict the liquid-phase volume distribution within the droplet and the temperature field distribution throughout the flow domain during its motion and ice formation process. As the droplet moves, it is cooled due to the temperature difference with its surrounding environment. When the droplet temperature drops to the freezing point (0 °C), ice formation commences. It can observe the process of freezing of droplets through  Fig.\ref{liquid  volume fraction feild}. In the initial stage of freezing, an ice shell is formed on the outer layer of the droplet, and then as the temperature inside the droplet continues to decrease, the ice layer continues to extend to the inside of the droplet until the droplet freezes completely. The droplets are not deformed during the freezing process, because the surface tension under the small Weber number ($We=0.01$) plays a major role in keeping the droplets from deforming \cite{wierzba1990deformation}. In addition, it can be seen that the freezing of the droplets is not uniform, and the tail part of the droplet freezes faster than the front part. While in Zhao and Dong's study \cite{zhao2017improved,zhao2017numerical}, the lower part freezes faster than the upper part during the droplet falling process, because the cooling effect of the lower part is stronger than that of the upper part. In our study, this phenomenon is due to the characteristics of the shear flow field. Based on the thermal field illustrated in Fig. \ref{Temperature feild}, it can be observed that the morphology of the temperature distribution within the droplet and its surrounding region evolves from a circular shape in the initial stage to a bullet-shaped configuration during the motion phase. This morphological transition aligns with the characteristics of the shear flow field. Furthermore, the internal temperature distribution of the droplet governs the resultant ice formation pattern within its interior, demonstrating consistency between the two aspects.

\subsection{Analysis of droplet freezing patterns and flow field characteristics}

In order to further explore the freezing patterns inside droplets, we present the volume fraction distribution contour and temperature distribution contour with different droplet Reynolds numbers and $D/W$ ratios with a Stefan number of 0.21 ($Ste=\frac{\mu_{\textbf{w}}{c_{p\textbf{w}}}}{l_\textbf{latent}}$), as shown in Fig.\ref{liquid volume fraction and temperature contour}. It finds that with the condition of $D/W=0.33$, both the ice formation and the temperature field distribution of droplet with $Re=1$ are more uniform, the tail and front of the droplet exhibit a symmetrical pattern, as shown in Fig.\ref{liquid volume fraction and temperature contour} (a); while for $Re=5$, the phase transition front of the droplet during freezing exhibits a pronounced anterior shift, leading to the gradual formation of a crescent-shaped unfrozen region. This region subsequently diminishes in size until complete solidification, at which point it fully disappears. Concurrently, the internal temperature field of the droplet develops a gradient-aligned crescent-shaped distribution, as visualized in Fig.\ref{liquid volume fraction and temperature contour}(b). The liquid volume fraction and temperature distribution of the droplet of $Re=6$ and $D/W=0.2$ shown in Fig.\ref{liquid volume fraction and temperature contour} (c) is uniform, with only a small shift in the unfrozen part. The internal freezing of the droplet of $Re=14$ and $D/W=0.47$ shown in Fig.\ref{liquid volume fraction and temperature contour} 
 (d) is uneven, and the freezing at the tail end is obviously faster than that at the front end, forming two peaks at 8 ms, and gradually forming a bow shape with the progress of freezing, until the droplets disappear when the droplets are completely frozen.

\begin{figure*}[t]
   \centering
    \includegraphics[width=16cm]{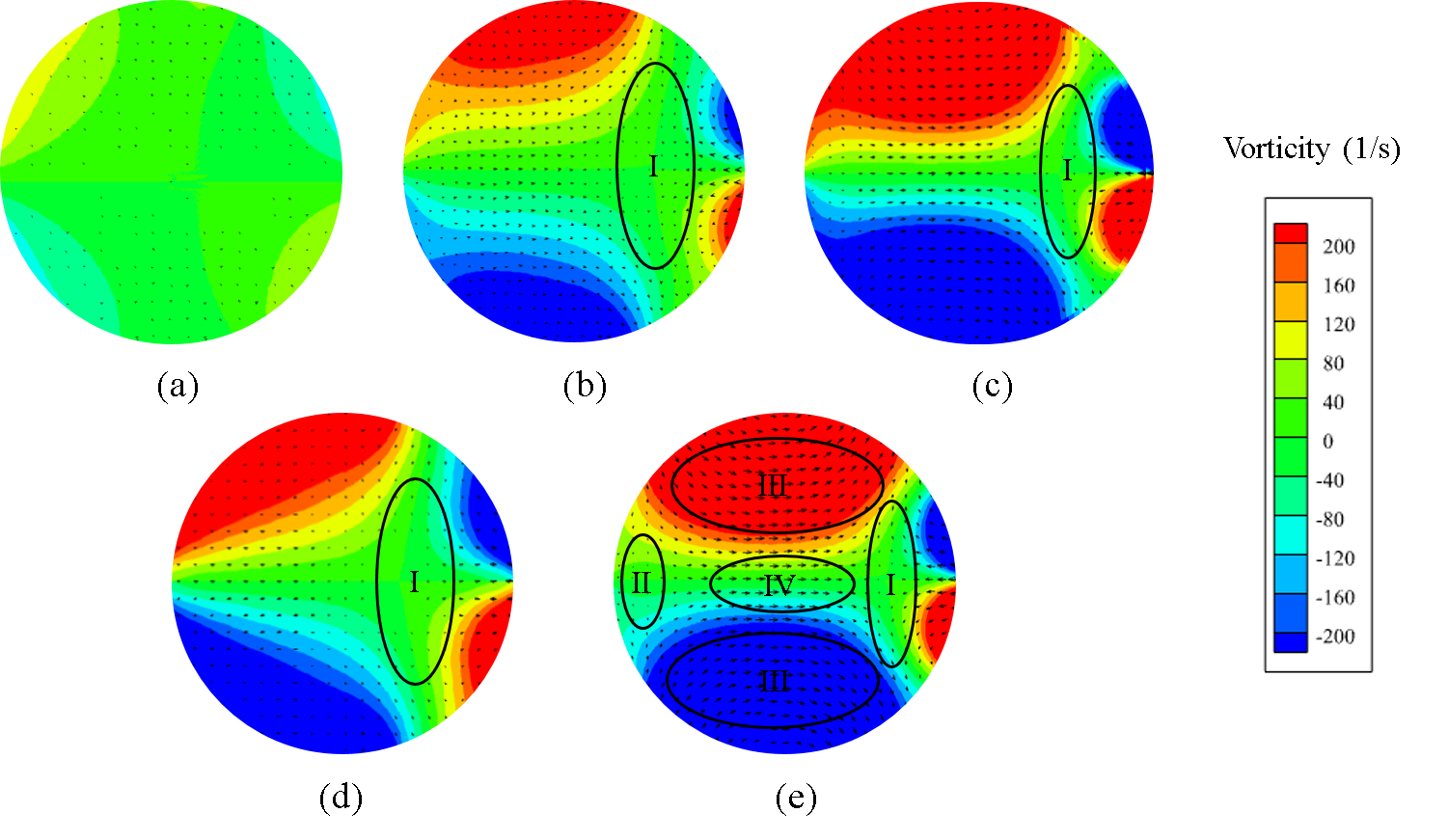}
    \caption{Internal flow patterns and vorticity contours of the droplet at t = 1 ms under various Reynolds and D/W ratios: (a) $Re=1$  and  $D/W=0.33 $; (b)$Re=5$ and $D/W=0.33$; (c) $Re=10$ and $D/W=0.33$; (d) $Re=6$ and $D/W=0.2$; (e) $Re=14$ and $D/W=0.47$.}
  \label{vector}
\end{figure*}

\begin{figure}[H]
   \centering
    \includegraphics[width=\linewidth]{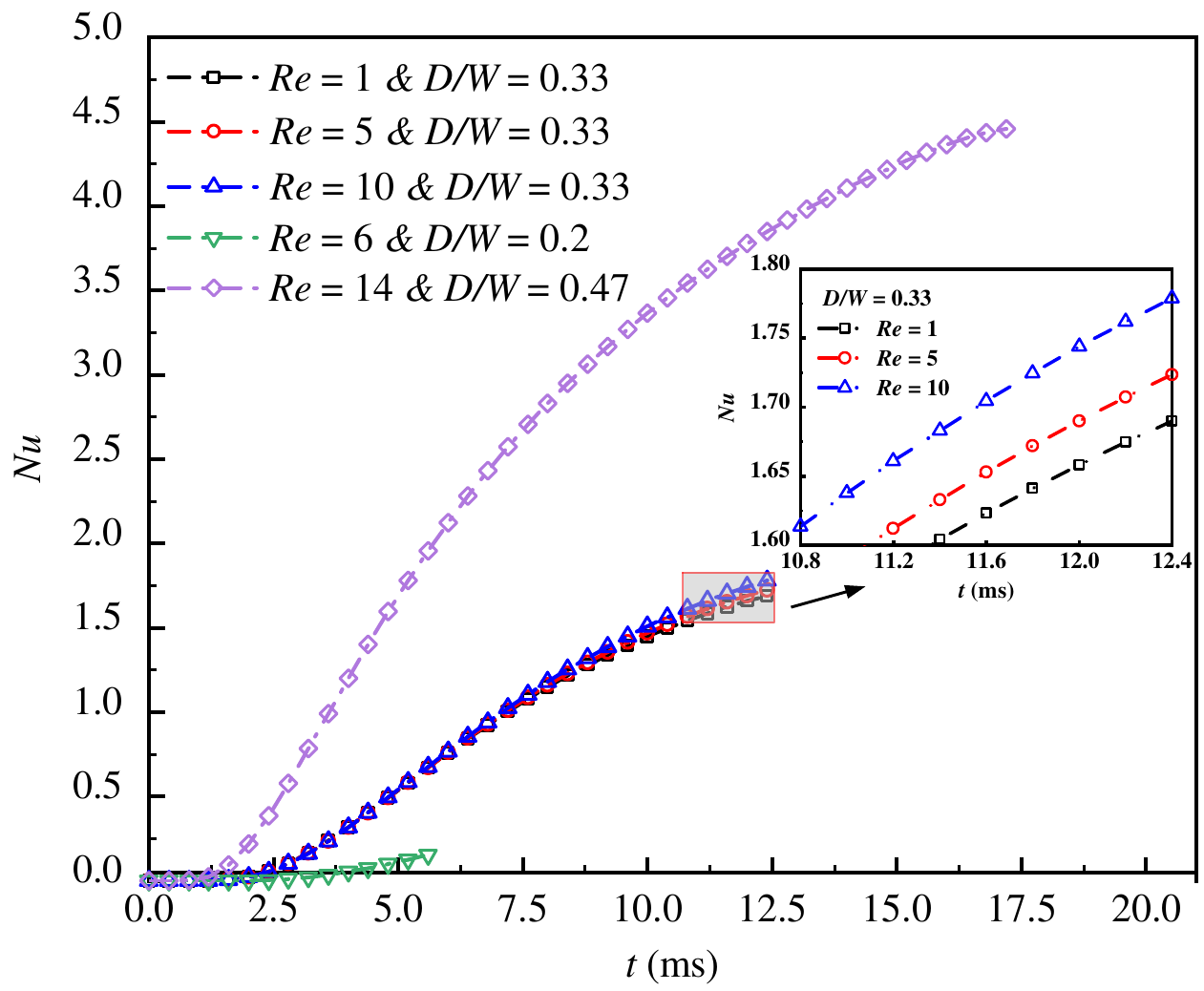}
    \caption{Temporal variation of the average Nusselt number ($Nu$) within the flow field across the five computational cases.}
  \label{Nu}
\end{figure}

To further elucidate the distinct ice formation pattens of the droplet, we isolate the transient flow features by subtracting the initial flow field, presenting solely the internal droplet motion and the vorticity contours within the droplet at t = 1 ms for the aforementioned five scenarios, as shown in Fig. \ref{vector}. As shown in Fig.\ref{vector} (a-c), for the case of $D/W=0.33$, the internal motion of the droplet intensifies with increasing Reynolds number. By comparing with the freezing interface evolution in Fig.\ref{liquid volume fraction and temperature contour} (a,b) and Fig.\ref{liquid volume fraction feild}, it is observed that the influence of Reynolds numbers on freezing primarily manifests in the droplet’s flow direction, while the spanwise freezing remains largely unaffected. Similarly, as illustrated in Fig.\ref{vector}  (c-e), the internal motion of the droplet becomes more vigorous with increasing $D/W$. Comparative analysis with Fig.\ref{liquid volume fraction and temperature contour} (c,d) and Fig.\ref{liquid volume fraction feild} reveals that larger $D/W$ ratios correlate with slower spanwise freezing, leading to non-uniform ice formation. Additionally, Fig.\ref{vector} (b-e) demonstrates the absence of convective motion in region I, where freezing predominantly occurs during the final stage. As shown in Fig.\ref{vector} (e), weaker motion in region II and intense motion in region III result in the distinct ice morphology at 4 ms. Following ice formation in region II, an accelerated freezing process in region IV further shapes the subsequent ice morphology. To further illustrate the impact of thermal convection and thermal conduction on freezing, based on formula:
\begin{equation}
    Nu=\frac{-\left.\frac{\partial T}{\partial n}\right|_{\text {wall }} \cdot W}{T_{\text{wall}}-T_{\text{ref}}}\label{Nu formulation}
    \end{equation}
 averaging by integration over the entire field yields the average Nusselt number ($Nu$) values of the flow field for the five cases, as shown in Figure \ref{Nu}. It can be observed that the Nusselt number is primarily influenced by droplet size. Larger droplets correspond to higher Nusselt numbers, indicating greater convective influence on droplet freezing. For $D/W=0.2$, small droplets freeze predominantly through heat conduction. When $D/W=0.33$ is held constant, variations in Reynolds number have minimal effect on Nusselt number. Simultaneously, Nusselt numbers less than two indicates that heat conduction dominates the freezing process under these conditions. At $D/W=0.47$, convection exerts a significant influence on droplet freezing.
 
\begin{figure*}[t]
    \centering
    \hfill
    \begin{minipage}{0.49\textwidth}
        \includegraphics[width=\linewidth]{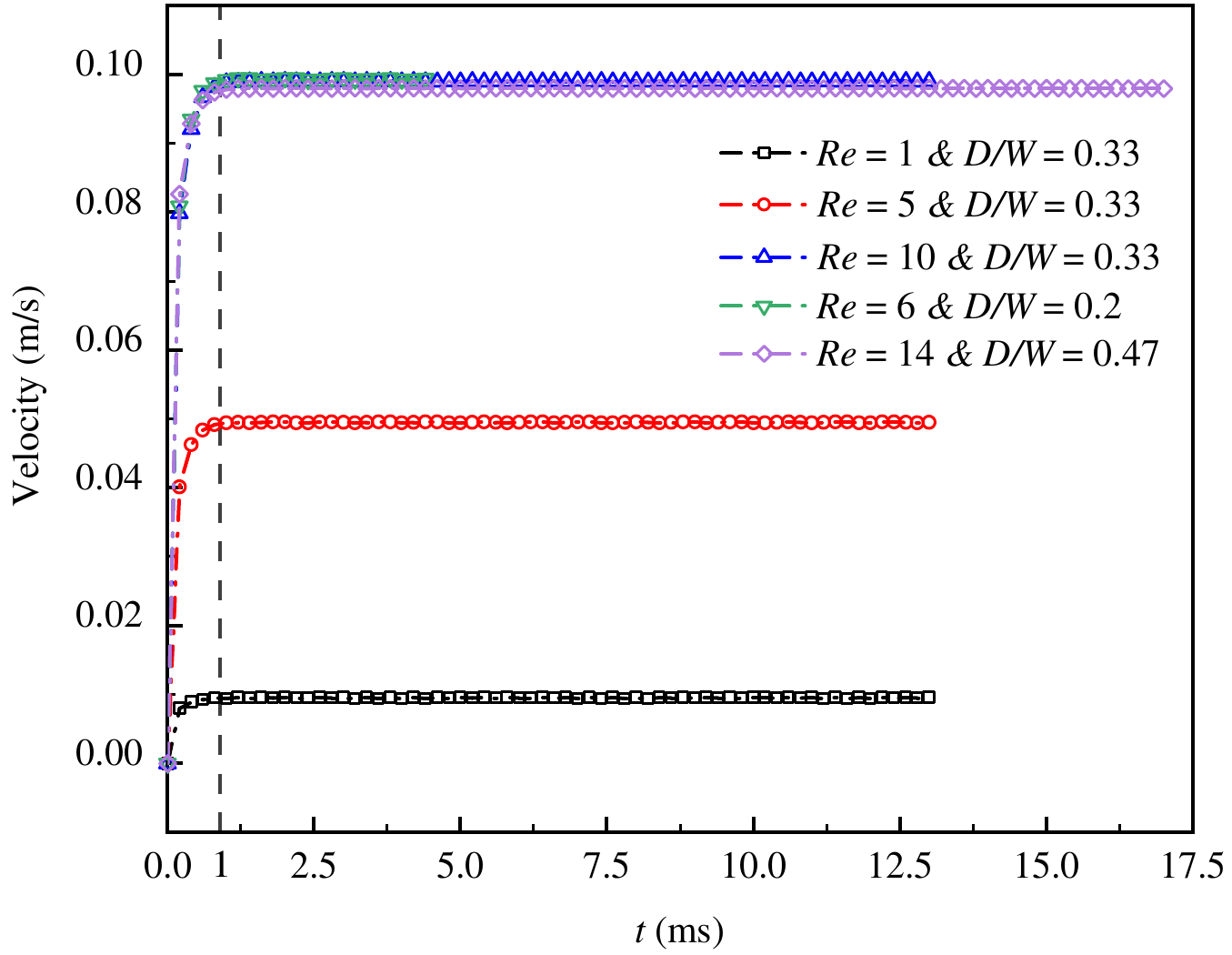}
        \caption*{(a)}
    \end{minipage}
    \begin{minipage}{0.49\textwidth}
        \includegraphics[width=\linewidth]{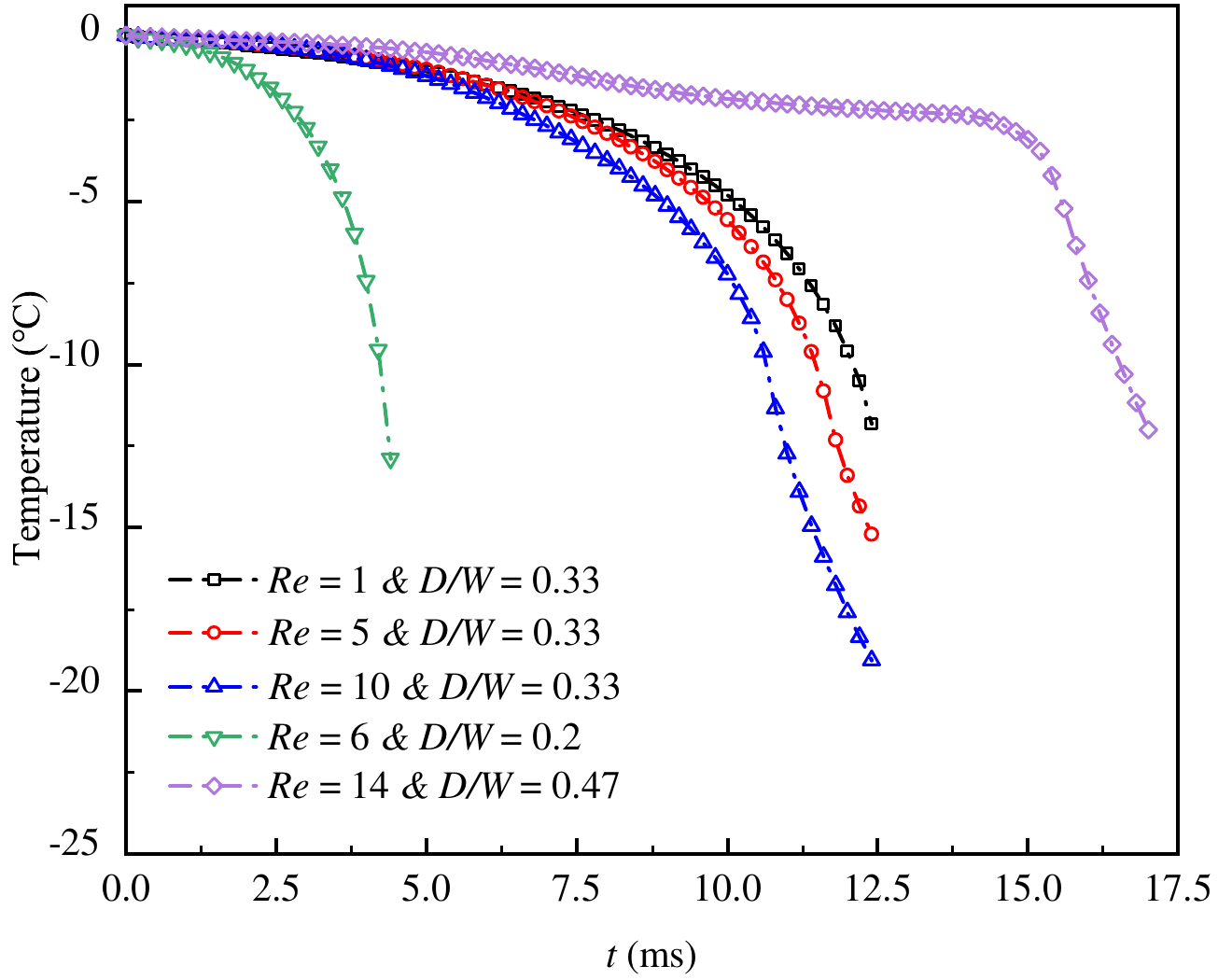}
        \caption*{(b)}
    \end{minipage}
    \hfill
    \begin{minipage}{0.49\textwidth}
        \includegraphics[width=\linewidth]{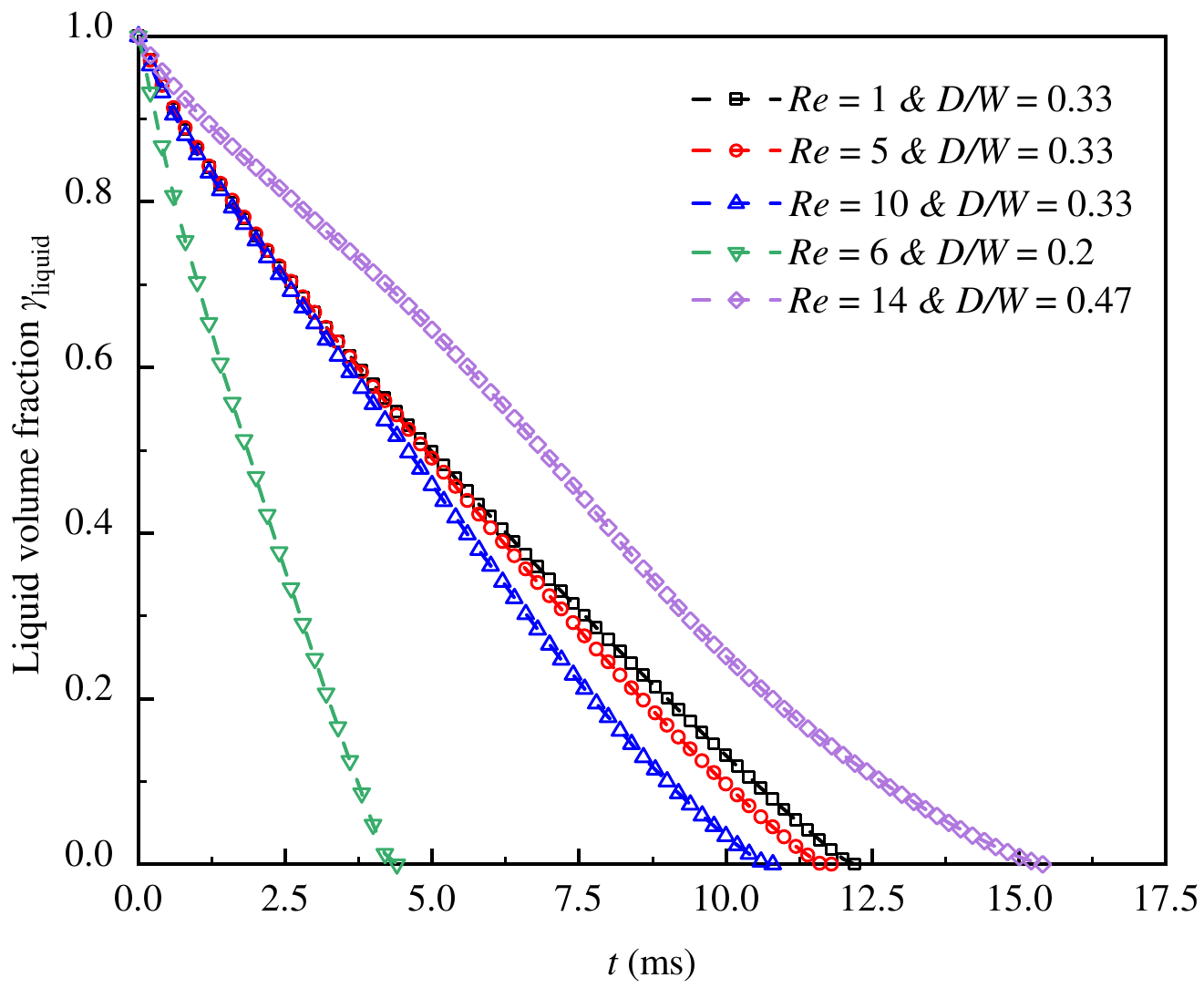}
        \caption*{(c)}
    \end{minipage}
    \begin{minipage}{0.49\textwidth}
        \centering
        \includegraphics[width=\linewidth]{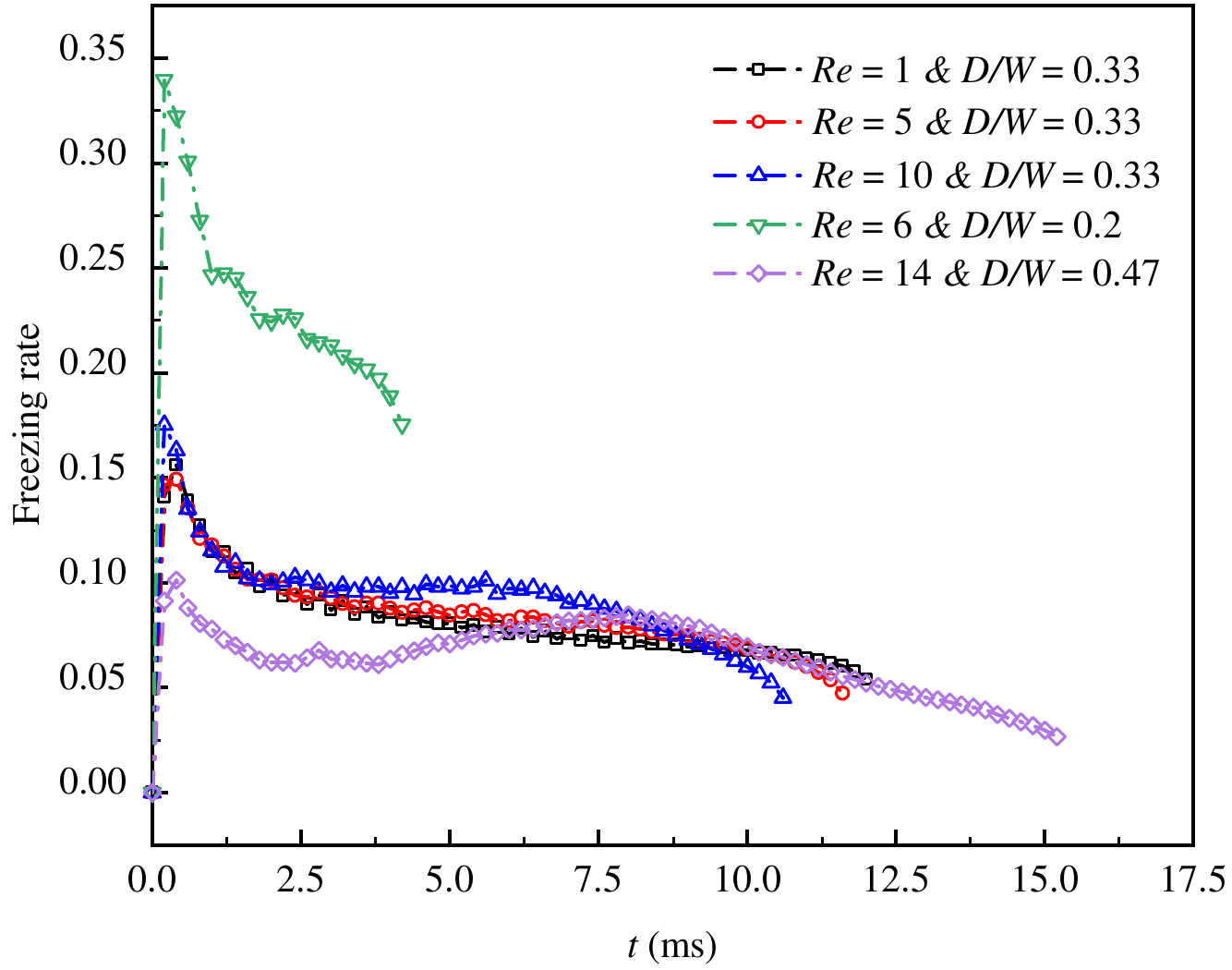}
        \caption*{(d)}
    \end{minipage}
    \caption{ velocity (a),  temperature (b),liquid volume fraction (c), and freezing rate (d) vary with time under the above five calculation conditions.}
    \label{Four}
\end{figure*}

Additionally, we record the temporal variations in velocity, temperature, liquid volume fraction and freezing rate of droplet under different Reynolds numbers and $D/W$ ratios, as illustrated in Fig.\ref{Four}. As shown in Fig.\ref{Four} (a), the droplet’s velocity stabilizes once the flow field reaches a steady state, indicating that freezing has negligible impact on the overall motion of the droplet in a stable flow feild. Fig.\ref{Four} (b) reveals that during the initial freezing stage, the temperature decreases gradually, but the cooling accelerates significantly toward the later phase. Once the droplet is fully frozen, the temperature drops rapidly. This behavior is attributed to the release of latent heat during freezing, combined with the enhanced thermal conductivity of the solid phase compared to the liquid state.From Fig.\ref{Four} (c) and (d), it is evident that larger $D/W$ ratios correspond to longer freezing times but slower freezing rates. As the Reynolds number increases, the freezing rate slightly rises, reflecting the influence of convective effects.

\subsection{Freezing time scaling relation}

\begin{figure*}[!t]
    \centering
    \begin{minipage}{0.49\textwidth}
        \includegraphics[width=\linewidth]{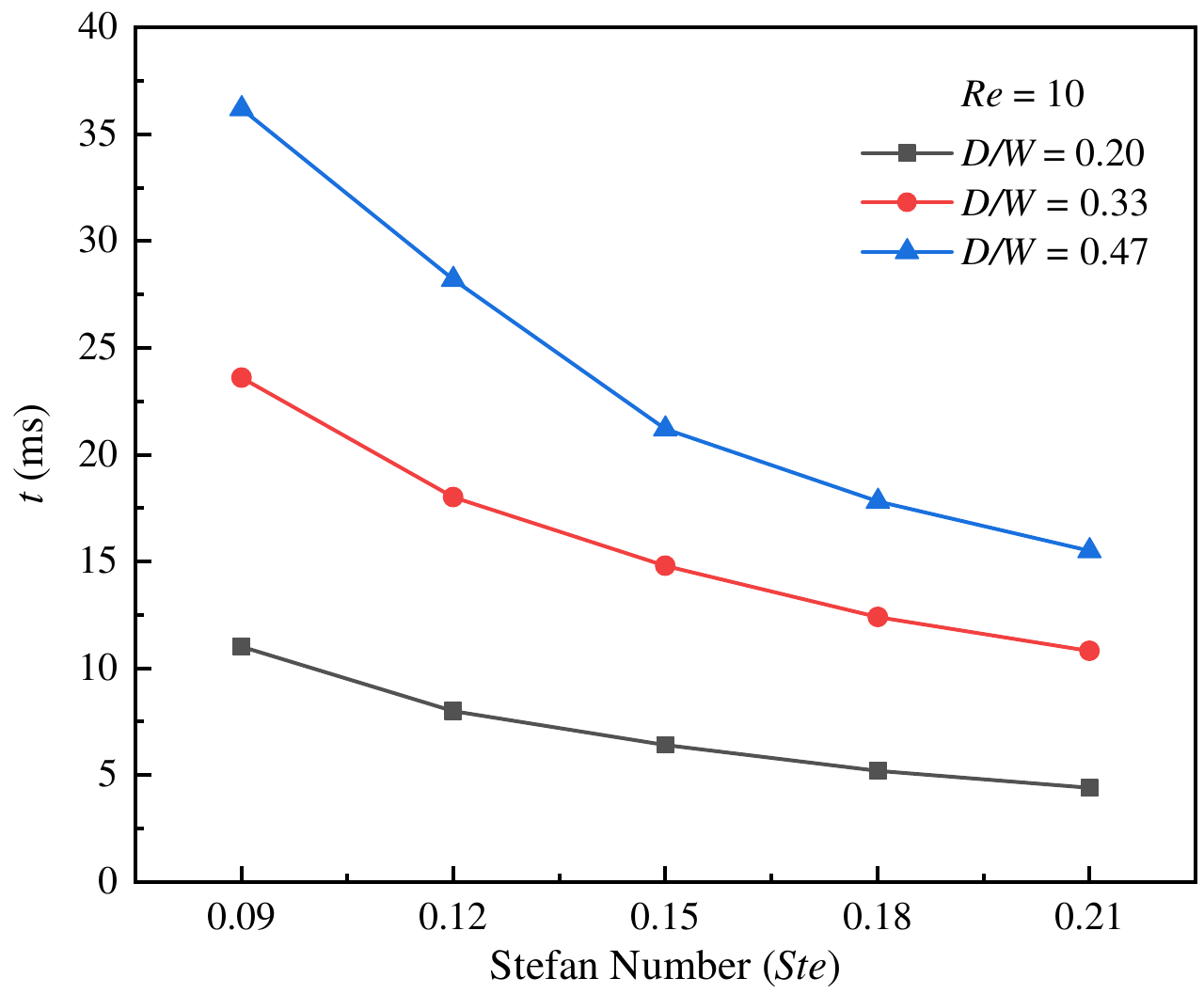}
        \caption*{(a)}
    \end{minipage}
    \begin{minipage}{0.49\textwidth}
        \includegraphics[width=\linewidth]{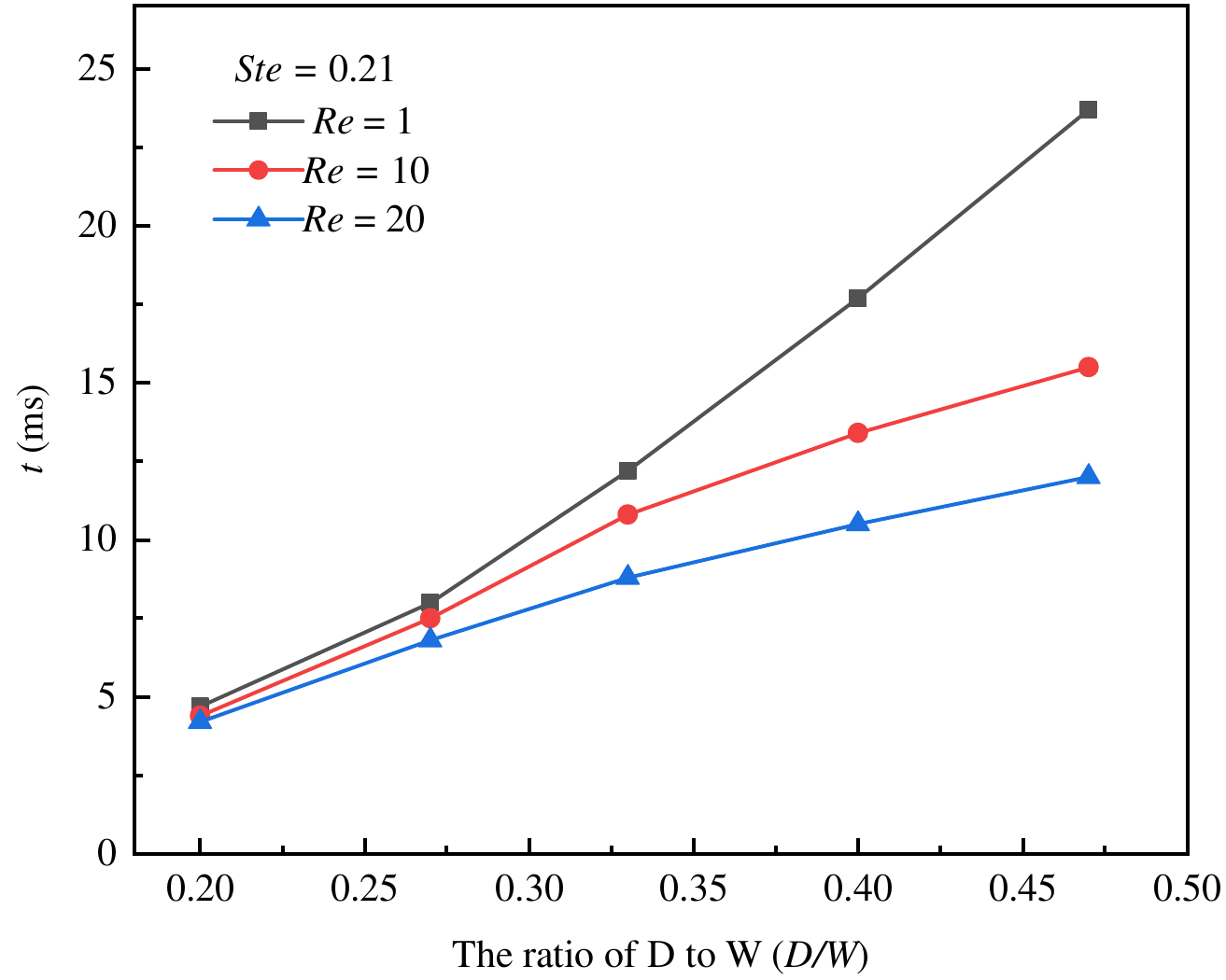}
        \caption*{(b)}
    \end{minipage}
    \begin{minipage}{0.49\textwidth}
        \centering
        \includegraphics[width=\linewidth]{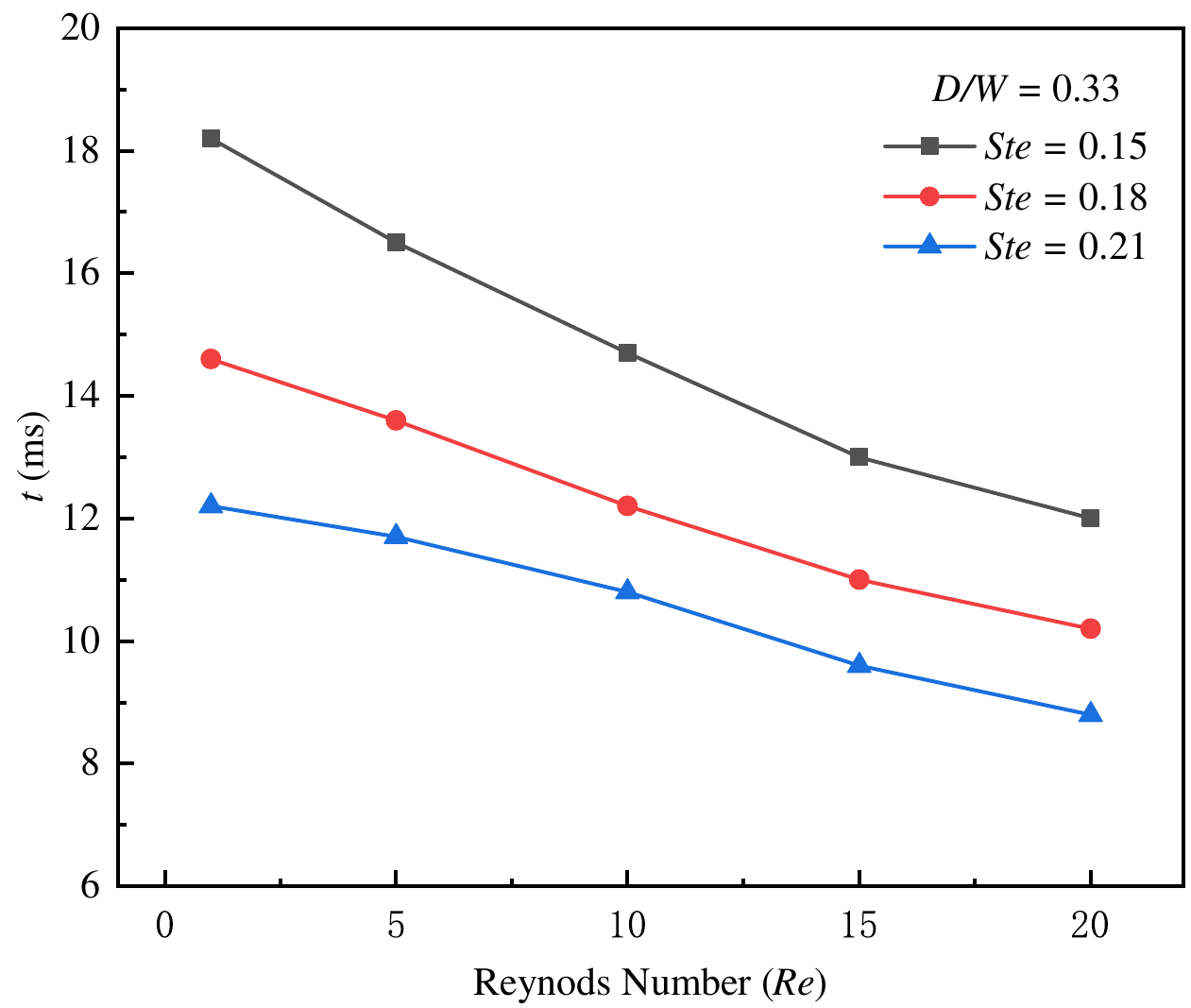}
        \caption*{(c)}
    \end{minipage}
    \caption{Effect of different dimensionless numbers on the freezing time of droplet:(a) Stefan number($Ste$), (b) The ratio of $D$ to $W$ ($D/W$) and (c) Reynolds number($Re$).}
    \label{Dimensionless numbers}
\end{figure*}

In order to explore the influence of factors in our system on the freezing time of droplets, we use dimensional analysis to express the freezing time and influencing parameters in the form of implicit functions, as follows:
\begin{equation}
    t_\text{final}=f(\rho _{\mathrm{w}},\mu_\text{{w}}, c_{p \mathrm{w}}, k_{\mathrm{w}}, l_\text{latent}, D,\Delta T,\rho _{\mathrm{o}},\mu_{\mathrm{o}}, c_{p\mathrm{o}}, k_{\mathrm{o}}, u_{\mathrm{max}}, W)\label{tf}
\end{equation}
Combining the dimensions of each parameter, we take $\rho_\mathrm{w}$, $u_{\mathrm{max}}$, $D$, and $c_{p\mathrm{w}}$ as the basic physical quantities to construct the following combination of dimensionless quantities:$\Pi_{1}=\frac{t_{\mathrm{f}}u_{\mathrm{max}}} D$, $\Pi_{2}=\frac{\mu_{\mathrm{w}}}{\rho_{\mathrm{w}} u_{\mathrm{max}} D}$, $\Pi_{3}=\frac{k_{\mathrm{w}}}{\rho_{\mathrm{w}} c_{p\mathrm{w}} u_\mathrm{max}D}$, $\Pi_{4}=\frac{l_\text{latent
}}{u_\mathrm{max}^{2}}$, $\Pi_{5}=\frac{c_{p\mathrm{w}}\Delta T }{u_\mathrm{max}^{2}}$, $\Pi_{6}=\frac{\rho_{o}}{\rho_\mathrm{w}}$, $\Pi_{7}=\frac{\mu_\mathrm{o}}{\rho_\mathrm{w}u_{\mathrm{max}} D}$, $\Pi_{8} =\frac{c_{p\mathrm{o}}}{c_{p\mathrm{w}}}$, $\Pi_{9}=\frac{k_{\mathrm{o}}}{\rho_{w} c_{p\mathrm{w} }u_\mathrm{max} D}$, $\Pi_{10}=\frac{W}{D}$. Then we obtain the well-known dimensionless quantity or the ratio of two contant parameters by combining the dimensions between the dimensions:$\pi_2^{*}=\Pi_{2}^{-1}=Re$, $\Pi_{3}^{*}=\Pi_{3}^{-1} \Pi_{2}=Pr$, $\Pi_{4}^{*}=\Pi_{4}^{-1}\Pi_{5}=Ste$, $\Pi_{7}^{*}=\Pi_{7} \Pi_{2}^{-1}=\frac{\mu_\mathrm{o}}{\mu_{w}}$, $\Pi_{9}^{*}=\Pi_{9} \Pi_{3}^{-1}=\frac{k_\mathrm{o}}{k_\mathrm{w}}$, $\Pi_{10}^{*}=\Pi_{10}^{-1}=\frac{D}{W}$.

In our studied system, we primarily investigate the influence of variable parameters $\Delta{T}$, $u_{max}$ and $D$ on the target parameter $t_\text{final}$, corresponding to the dimensionless numbers $Ste$, $Re$ and $D/W$. The relationships between these dimensionless numbers and freezing time are illustrated in Fig.\ref{Dimensionless numbers}. A clear negative correlation is observed between the Stefan number and freezing time from Fig.\ref{Dimensionless numbers} (a), as well as between the Reynolds number and freezing time from Fig.\ref{Dimensionless numbers} (b). Specifically, a higher Stefan number corresponds to a lower oil-phase temperature, enabling the droplet to absorb more cooling energy during freezing, thereby shortening the freezing time. Similarly, a higher Reynolds number enhances convective heat transfer, accelerating the freezing process and reducing freezing time. In contrast, the $D/W$ ratio exhibits a positive correlation with freezing time, as larger droplets require greater heat removal to freeze. At $Re = 1$, freezing time follows an approximate exponential relationship with $D/W$, this relationship diminishes at $Re = 10$ and $Re = 20$, further highlighting the influence of convective heat transfer on the freezing process. By fitting the data of Fig.\ref{Dimensionless numbers} (a-c) graphs, a formula is obtained that relates the $t_\text{final}$ with the $Ste$, {$Re$} and ${D/W}$,  as follows:

\begin{equation}
    t_\text{final}\sim a_{1}{Ste}^{a_{2}} {Re}^{a_{3}} {(D/W)}^{a_{4}}\label{fit}
\end{equation}

\begin{table*}[ht]
\centering
\caption{Comparison of Expressions for Droplet Freezing Time and Influencing Factors under Different Simulation Conditions}
\resizebox{\textwidth}{!}{ 
\begin{tabular}{@{}llccl@{}}
\toprule
\multirow{2}{*}{Simulation conditions} & \multirow{2}{*}{Freezing time expression} & \multicolumn{2}{c}{Influencing factor index}  \\ 
\cmidrule(lr){3-4}
 & & $D$ & $\Delta T$ & \\ 
\midrule
Freezing of static liquid droplets on a cold plate\cite{zhang2018simulation} & 
$\begin{aligned}
t_{\text{final}} = 0.0534 &V^{0.597}\theta^{1.6878} \\
&\times \Delta T^{-1.4078} + 2.2682  
\end{aligned}$ & 
1.791 & -1.4078  \\
\addlinespace

Freezing of a free falling droplet in cold environments\cite{sultana2017phase} & 
$\begin{aligned}
t \approx \frac{R^{2}}{\alpha_{\mathrm{eff}}} \bigg(1.3 + \frac{4.68}{\text{SteNu}}\bigg)
\end{aligned}$ & 
2 & -1 \\
\addlinespace

Our study & 
$\begin{aligned}
t_{\text{final}} \sim a_{1}\text{Ste}^{a_{2}} \text{Re}^{a_{3}} (D/W)^{a_{4}}
\end{aligned}$ & 
1.4185 & -0.9074 \\ 
\bottomrule
\end{tabular}
}
\label{tab:3}
\end{table*}

The influence of other fixed dimensionless numbers, including the droplet Prandtl number ($Pr$), is incorporated in the coefficient $a_1$. The fitting obtain the constants $a_1=18.02$, $a_2=-0.91$, $a_3=-0.12$ and $a_4=1.42$. The fitted exponents show that $|a_4| > |a_2| > |a_3|$, indicating that droplet size and temperature have a more significant impact on droplet frzeeing. In comparison with the work of the other two researchers, Zhang \cite{zhang2018simulation} and Sultana \cite{sultana2017phase}, as shown in Table \ref{tab:3}, the expression for droplet freezing time highlights the droplet size and temperature as significant influencing factors. We compare their exponents. For the size effect, we consistently express it in terms of droplet diameter ($D$). The exponent comparison $1.4185 < 1.791 < 2$ shows that the sensitivity to droplet size is less pronounced in our work than in the others. This indicates that the confined space suppresses the sensitivity to droplet size, meaning that changes in droplet size have a relatively lower impact on freezing time. For the temperature effect, we use the temperature difference between the droplet and the environment ($\Delta T$). The comparison $|-0.90741| < |-1| < |-1.4078|$ similarly reveals that the temperature sensitivity is lowest in our work. This suggests that the flow enhances heat transfer, diminishing the dominant role of temperature observed in pure conduction scenarios. However, due to the influence of flow, convective heat transfer overall accelerates the droplet freezing rate.

\begin{figure}[H]
   \centering
   \vspace{-10pt}
    \includegraphics[width=0.9\linewidth]{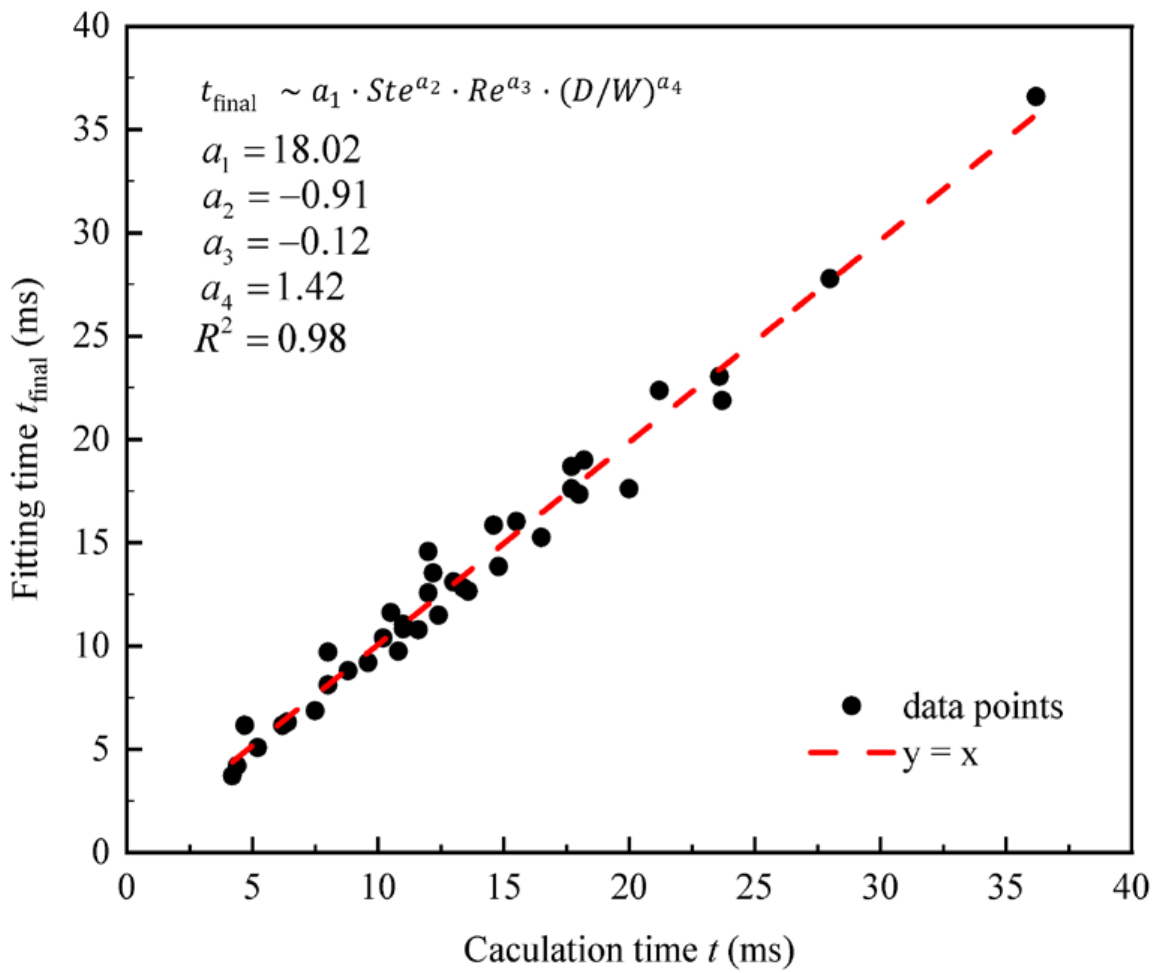}
    \caption{Comparion of Simulation calculation of freezing time and fitting freezing time}
  \label{fitting}
\end{figure}

\section{Summary and conclusions}\label{Conclusion}

This study systematically investigates the freezing process of the moving droplet in microchannels, with the following main conclusions:

Under supercooled conditions, the freezing of the moving droplet exhibits a surface-initiated solidification process, where the freezing front progressively advances inward until complete solidification is achieved. Shear flow effects induce significant spatial non-uniformity in the freezing process, with accelerated solidification observed in the tail region compared to the leading edge. The resultant ice morphology is governed by the coupled influence of the Reynolds number and $D/W$ ratio, where the Reynolds number predominantly determines the streamwise ice formation while the $D/W$ ratio controls spanwise development. At low Reynolds numbers, suppressed internal convection leads to symmetric radial freezing patterns, whereas higher Reynolds numbers enhance convective heat transfer from the droplet interior to the surface, causing forward displacement of the freezing front and asymmetric solidification. Furthermore, while smaller $D/W$ ratios promote uniform ice formation, larger ratios induce spanwise freezing stagnation and increased front-edge irregularity, ultimately producing a distinctive arc-shaped freezing pattern. Additionally, the release of latent heat during phase change causes a gradual increase in the droplet’s cooling rate. Post-freezing, the solidified portion exhibits higher thermal conductivity than the liquid phase, leading to a rapid temperature drop once full solidification is achieved. And under steady flow conditions, the droplet’s velocity stabilizes and remains nearly constant throughout the freezing process.

Through dimensional analysis, the nonlinear influence of the Stefan number, Reynolds number, $D/W$ ratio on freezing time is revealed, and a scaling relation for freezing time is derived:$t_\text{final}\sim18.02 {Ste}^{-0.91}{ Re }^{-0.12} (D/W)^{1.42}$. The Stefan and Reynolds numbers exhibit negative correlations with freezing time, where higher Stefan numbers and Reynolds number numbers shorten freezing duration, while $D/W$ ratio shows a positive correlation, with larger $D/W$ ratios prolonging freezing time. The fitted coefficients indicate that droplet size and temperature exert more dominant effects on droplet freezing. 

The findings advance the understanding of phase-change processes during microscale droplet freezing under motion and provide a theoretical foundation for designing microfluidic systems operating in low-temperature environments or scenarios prone to droplet freezing.

\section*{Declaration of Competing Interest}
The authors declare that they have no known competing financial interests or personal relationships that could have appeared to influence the work reported in this paper.

\newpage
\bibliographystyle{elsarticle-num-names} 
\bibliography{rev}

\appendix

\section{Validation of the Stefan problem}
\label{stefan problem}
\begin{figure}[H]
   \centering
    \includegraphics[width=\linewidth]{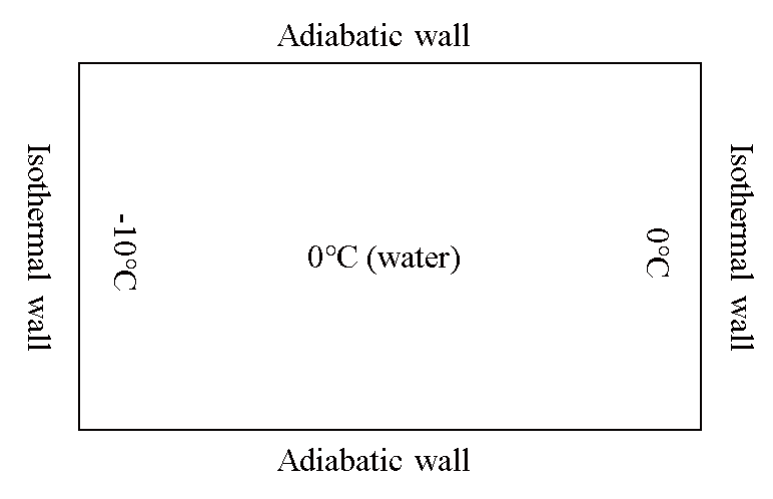}
    \caption{Boundary conditions and initial interface position for Stefan problem}
  \label{Stefan1}
\end{figure}

Researchers usually use the Stefan problem to prove the reliability of heat and mass transfer algorithms in phase transition simulations \cite{welch2000volume,dallaire2017numerical}. In this subsection, the reliability of the model is verified by calculating a single-phase Stefan question for the solidification of water in a saturated state at 0°C. As shown in Fig.\ref{Stefan1}, a physical model of the single-phase Stefan problem is established. Initially, the space is filled with saturated water with a temperature of $T_{\text{sat}}=0$°C, the left wall is kept at a constant temperature of $T_{\text{wall}}=-10$°C, the right wall is kept at the same temperature as the saturation temperature, and the upper and lower walls are insulated. For calculation, the density ($\rho_{\mathrm{w}}=998.2\mathrm{kg} / \mathrm{m}^{3}$), the specific heat ($c_{p\mathrm{w}}=4182\mathrm{J}/\left(\mathrm{kg} \cdot \mathrm{K}\right)$), the thermal conductivity ($k_{\mathrm{w}}=0.6\mathrm{W} /(\mathrm{m} \cdot \mathrm{K})$), and the latent heat  
 ($h_{\mathrm{w}}=333146 \mathrm{J} / \mathrm{kg}$) of water are used.
The solution to the problem is as follows\cite{alexiades2018mathematical}:
\begin{equation}
     \delta\left( t \right) = 2\lambda \sqrt {\alpha t}
\end{equation}
\begin{equation}
    T\left( {x,t} \right) = {T _{{\rm{wall  \;}}}} + \left( {\frac{{{T _{{\rm{wall\;}}}} - {T_{{\rm{sat\;}}}}}}{{{\rm{erf}}\left( \lambda  \right)}}} \right){\rm{erf}}\left( {\frac{x}{{2\sqrt {\alpha t} }}} \right)
\end{equation}
where $\alpha=k_{\text{w}}/{\rho_{\textbf{w}} c_{p}}$ is the diffusion coefficient, $\operatorname{erf}(\lambda)$ is the error function and $\lambda$ is the solution of the transcendent equation.
\begin{equation}
    \lambda \exp \left(\lambda^{2}\right) \operatorname{erf}(\lambda)=\frac{c_{p\text{w}}\left(T_{\text {wall }}-T_{\text {sat }}\right)}{h_{\text{w}} \sqrt{\pi}}
\end{equation}

\begin{figure}[H]
    \centering
    \includegraphics[width=\linewidth]{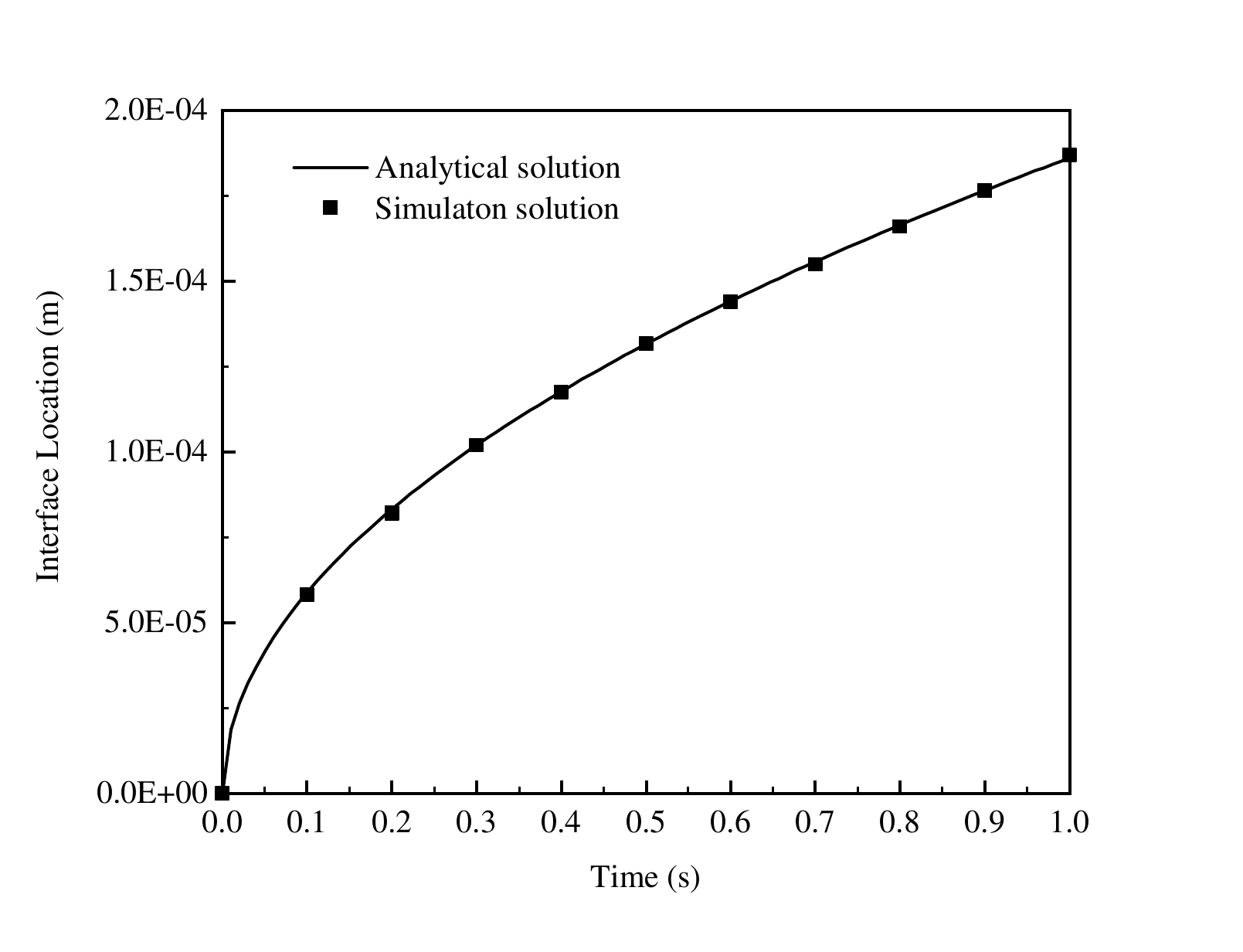}
    \caption{Comparison of simulated and theoretical interface positions in 0 to 1 s}
    \label{stefan2}
\end{figure}

The comparison of the analytical solution and the simulated solution is shown by Fig.\ref{stefan2}, the position of the interface is calculated as a function of time, and the position change of the phase interface in time of 1s can be seen, the results of the numerical simulation are consistent with the results of the theoretical analytical solution, indicating that the simulation of the Stefan problem is reliable.

\end{document}